# Surface phonon-polaritons enhance thermal conduction in SiN nanomembranes

Y. Wu[1,*], J. Ordonez-Miranda[2,†], S. Gluchko[1], R. Anufriev[1], S. Volz[1,3], and M. Nomura[1]

[1]Institute of Industrial Science, The University of Tokyo, Tokyo 153-8505, Japan
[2]Institut Pprime, CNRS, Université de Poitiers, ISAE-ENSMA, F-86962, Futuroscope Chasseneuil, France
[3]Laboratory for Integrated Micro Mechatronic Systems/National Center for Scientific Research-Institute of Industrial Science (LIMMS/CNRS-IIS), The University of Tokyo, Tokyo 153-8505, Japan
*yunhui@iis.u-tokyo.ac.jp
†jose.ordonez@cnrs.pprime.fr

## ABSTRACT

Surface phonon-polaritons can carry energy on the surface of dielectric films and thus expected to contribute to heat conduction. However, the contribution of surface phonon-polaritons (SPhPs) to thermal transport has not been experimentally demonstrated yet. In this work, we experimentally measure the effective in-plane thermal conductivity of amorphous silicon nitride membrane and show that it can indeed be increased by SPhPs significantly when the membrane thickness scales down. In particular, by heating up a thin membrane (<100 nm) from 300 to 800 K, the thermal conductivity increases twice due to SPhPs contribution.

## Introduction

Thermal conduction becomes less efficient as structures scale down since phonon-boundary scattering becomes predominant. Therefore, thermal management becomes more challenging in micro-electronic or optical devices[1–3]. Over the past decade, significant research efforts have been devoted to the study of surface waves, due to the fact that the surface effects predominate over the volumetric ones with the high surface-to-volume ratio in nanomaterials. Moreover, the surface waves potentially serve as novel heat carriers[4, 5] and emerge considerable potential applications to improve the thermal performance and stability of nanoscale devices in electronics[6–8]. Recent studies aim at revealing a novel heat carrier, i.e. the phonon polariton, which might be able to overcome the decrease of conventional thermal transport and even lead to its enhancement due to a surface effect.

Surface phonon-polaritons (SPhPs) are the evanescent electromagnetic waves generated by the hybridization of the evanescent electromagnetic surface waves and optical phonons, propagate along the interface of polar medium (such as $SiO_2$ or SiC) and a dielectric[9–11]. The mean free path of this evanescent wave is predicted to be in the range of hundreds of micrometers[12], which is significantly larger than the mean free path of the acoustic phonons. The SPhPs contribution to in-plane thermal conductivity of a nanomembrane is analytically investigated based on the Maxwell equations and the Boltzmann transport equation. They found that for a 40-nm-thick $SiO_2$ membrane heated up to 500 K, the total thermal conductivity increases nearly 100% over the intrinsic phonon thermal conductivity due to the SPhPs contribution[9, 10].

Moreover, silicon nitride (SiN) thin membranes have extensive applications in modern microelectronic and optoelectronic industries. Amorphous SiN is commonly used as dielectric material in MEMS devices, whose micro-heater and thermal sensors are thermally isolated in suspended thin membranes. The thermal performance of these SiN-based structures is strongly determined by their in-plane thermal conductivity[13], whose temperature variations also allow reducing heat losses in the quantum regime of heat conduction[14]. The phonon thermal conductivity of a thin membrane generally decreases as its thickness reduces to nanoscales due to the predominant phonon-boundary scattering[15]. This reduction of the thermal performance has dramatic engineering effects, such as the overheating, low reliability, and reduced lifetime of electronic components. The in-plane thermal conductivity of amorphous SiN membranes at room temperature have been measured by Griffin *et al.*[16], who reported values around 2 $Wm^{-1}K^{-1}$, for thicknesses ranging from 60 nm to 8.5 $\mu$m. This thickness-independent behavior was also observed by Mastrangelo *et al.*[17] for amorphous SiN membranes with thicknesses between 2 and 5 $\mu$m, whose average in-plane thermal conductivity was 3.2 $Wm^{-1}K^{-1}$. By contrast, Lee *et al.*[18] reported a reduction of the in-plane thermal conductivity as the SiN membrane thickness reduces through values within the interval (20 - 260 nm), such that its values become independent of the membrane thickness for thicknesses greater than 200 nm. All these measurements for the in-plane thermal conductivity of amorphous SiN membranes were done at room temperature and there is no currently experimental evidence of its values at higher temperature, to the best of our knowledge. These latter high-temperature values are expected to increase with temperature due to the expected energy contribution of SPhPs propagating along SiN nanomembranes.

In this work, we aim at revealing the existence of SPhPs and the contribution in thermal conduction in nanomembranes above room temperature. We conduct a synchronized excitation-detection experiment with a micro time-domain thermoreflectance ($\mu$TDTR) setup on a SiN suspended membrane to study the contribution of SPhPs to in-plane heat conduction. Our interest focuses on above room temperatures ranging from 300 to 800 K, where conventionally expect the material becomes less conductive due to the Umklapp scattering of phonons.

## Materials and Methods

### Sample preparation

Samples of amorphous SiN membranes are suspended in $1.0 \times 1.0$ mm$^2$ rectangular windows and with thicknesses of 30, 50, 100, and 200 nm. These high stress ($\approx$ 250 MPa) membranes were flat (curvature radius of 4 m), did not exhibit wrinkles and had an average roughness of 0.5 nm, as shown in Fig. 1. The deposition of the circular Al pad of 5 $\mu$m in diameter and 70 nm in height, on top of each membrane was done by means of electron beam lithography and electron beam assisted physical deposition (Ulvac EX-300) processes. The separation distance between these metallic pads was chosen large enough ($> 200$ $\mu$m) to minimize their contribution to the total thermal properties of the SiN membranes. After the lift-off process, an optical microscopy image is taken, as shown in Fig. 1(a), to allow the visualization of the quality of the membrane after the fabrication process. In addition, an AFM image is taken, as shown in Fig. 1(b), to characterize the surface roughness of the SiN membrane. We estimated the surface roughness to less than 1 nm, which should not influence SPhPs transport.

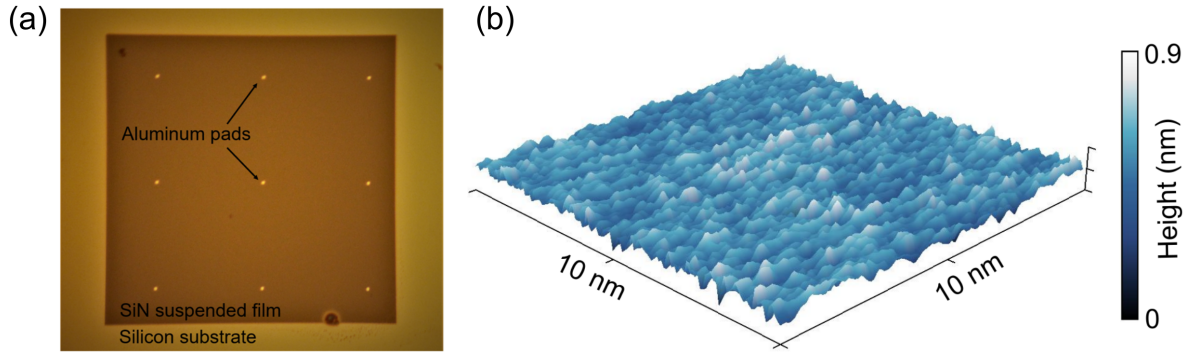

**Figure 1.** (a) Microscope image of the top view of a SiN membrane after fabrication; (b) AFM image of a SiN sample with the peaks height are smaller than or equal to 0.9 nm, which yields a relatively small roughness of 0.5 nm.

### Experimental setup

To study heat conduction at different temperatures, we placed the samples in a vacuum chamber with a temperature controlled stage for high (300 - 800 K) temperature measurement. To avoid the convection, the pressure in the chamber was below $10^{-3}$ Pa.

To measure the thermal conductivity of the membranes, we used micro time-domain thermoreflectance method ($\mu$TDTR), schematically shown in Fig. 2. In this all-optical pump-probe technique, a continues-wave "probe" (785 nm) and a pulsed "pump" (642 nm) laser beams are focused on the aluminum pad by an ($\times$50) microscope objective. The probe beam is used to continuously measure the reflectance of the aluminum pads by monitoring the intensity of the reflected beam in the photodetector plugged in a digital oscilloscope. The pump beam is used to periodically heat up the aluminum pad, which changes its reflectance coefficient. Since the reflectance is proportional to the temperature *via* the thermoreflectance coefficient, each pulse of the pump laser generates a jump in the signal of the probe laser. As heat gradually spreads from the aluminum pad through the underlying membrane, the temperature and thus the reflectance of the pad return to the initial values. This process is measured by the probe laser as a gradual return of the reflected laser intensity.

During one iteration, the TDTR system integrates the data from the lock-in detector for $10^4$ pump pulses. To further reduce the nose, $10^3$ of the such iterations are averaged to obtain one decay curve. To extract the thermal conductivity from the measured cures, we used an analytical model based on heat diffusion equation and Laplace transform.

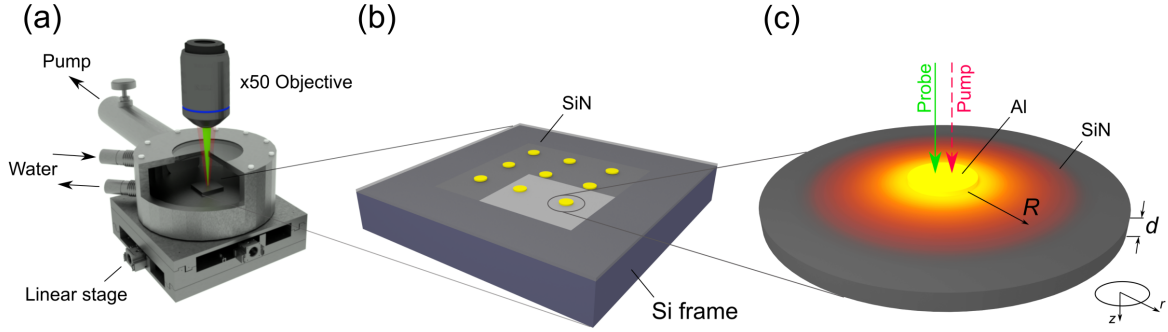

**Figure 2.** Scheme of the TDTR experimental setup: (a) a continuous probe laser and a pulsed pump laser are focused by an (×50) microscope objective; (b) an image of suspended SiN membrane supported by Si frame, which has been placed in a water-flow heating chamber; (c) a pump laser beam periodically heats up a circular aluminum pad, while the pad reflectance is continuously monitored via the reflected intensity of a probe laser beam impinging into a photodetector. The localization of the aluminum pad is done by means of a lamp and a camera.

## Results and Discussion

The in-plane thermal conductivity of four different thin membranes ($d$ = 30, 50, 100 and 200 nm thick) have been measured versus temperature from 300 to 800 K. The pulse of the pump laser that we applied here is 100$\mu$s. We heated up the samples by 20 K steps and, for each temperature, let the samples reach thermal equilibrium (a typical 10 minutes delay was noted). The temperature increased due to the probe laser is within 20 K.

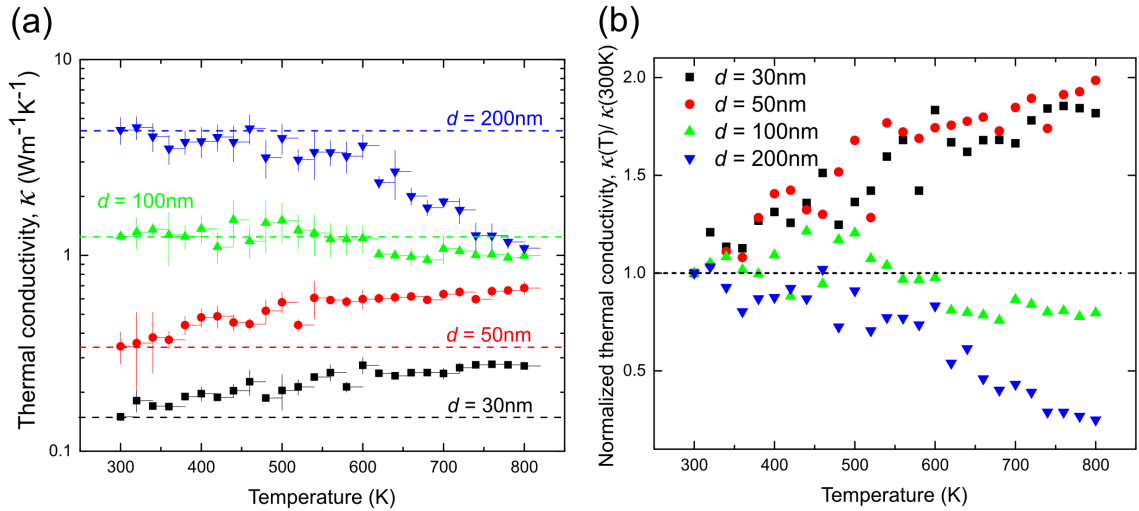

**Figure 3.** Temperature variations of the (a) absolute and (b) relative values of the in-plane thermal conductivity retrieved from the TDTR signals of SiN membranes with four thicknesses, where the dashed line is the guide for the eye.

The experimental data of the in-plane thermal conductivity for various membrane thicknesses of amorphous SiN are presented in Fig. 3 (a). For the 200 nm membrane, we obtained the thermal conductivity of 4.36 ± 0.69 Wm$^{-1}$K$^{-1}$ at 300 K. Up to 600 K, the thermal conductivity is nearly temperature-independent. Above 600 K, the thermal conductivity is inversely proportional to temperature and the slope is approximately $1/T^2$. This is due to the fact that Umklapp scattering. In this thickness, the SPhPs modes at the two surfaces are not coupled.

For the 100 nm membrane, the thermal conductivity is nearly independent of temperature. We obtained 1.25 ± 0.078 Wm$^{-1}$K$^{-1}$ and 0.99 ± 0.036 Wm$^{-1}$K$^{-1}$ at 300 K and 800 K, respectively. The decrease of thermal conductivity at high temperature due to Umklapp scattering is potentially compensated by the contribution of SPhPs.

The stronger thermal conductivity enhancement at high temperature is observed on the thinnest films, as expected from the contribution of the SPhPs. For the 50 nm membrane, we obtained the thermal conductivity of 1.25 ±0.078 Wm$^{-1}$K$^{-1}$ and $0.99 \pm 0.036$ Wm$^{-1}$K$^{-1}$ at 300 K and 800 K, respectively. Similar enhancement in thermal conductivity is observed in 30 nm membrane, where we obtained the thermal conductivity of $1.25 \pm 0.078$ Wm$^{-1}$K$^{-1}$ and $0.99 \pm 0.036$ Wm$^{-1}$K$^{-1}$ at 300 K and 800 K, respectively.

We normalized the values for different temperature by the ones at room temperature, shown in Fig. 3 (b). For thinner membranes, the SiN membrane becomes twice more conductive when it is heated up to 800 K. In this regime of thickness, the membrane is thin enough for the interaction between the SPhPs generated on either side of the membrane.

## Conclusions

We demonstrated that thermal conductivity was enhanced by SPhPs in amorphous SiN membrane above room temperature. In membranes thinner than 100 nm we observed the heat conduction by SPhPs, which caused an increase of the thermal conductivity with temperature. The results presented here point to overcome the reduction of thermal conductivity due to the size effect or even enhance total ability of heat conduction when the structure scales down. This phenomenon directs the future applications in the field of heat transfer, thermal management, near-field radiation and polaritonics.